\newcommand{\etal}{et al.}
\documentclass{aa}
\usepackage{times}
\usepackage{amssymb}
\usepackage{graphicx}
\begin{document}

   \thesaurus{03     
              (13.07.1;   
               11.19.3;  
               09.04.1)} 
\title{GRB 970228 and GRB 980329 and the Nature of Their Host Galaxies}

\author{D. Q. Lamb 
   \and Francisco J. Castander
   \and Daniel E. Reichart}
\institute{Department of Astronomy and Astrophysics, University
of Chicago, \\ 5640 South Ellis Avenue, Chicago, IL 60637}

\date{Received December 15, 1998}

\maketitle

\begin{abstract}
We find that the local galactic extinction towards the field of
gamma-ray burst GRB970228 is $A_V=1.09^{+0.10}_{-0.20}$, which implies
a substantial dimming and change in the spectral slope of the intrinsic
GRB970228 afterglow.  We measure a color  $(V_{606}-I_{814})_{ST} =
-0.18^{+0.51}_{-0.61}$ for the extended source coincident with the
afterglow.  Taking into account our measurement of the extinction
toward this field, this color implies that the extended source is most
likely a galaxy undergoing star formation, in agreement with our
earlier conclusion (\cite{CL98}). In a separate analysis, we find that
the inferred intrinsic spectrum of the GRB 980329 afterglow is
consistent with the predictions of the simplest relativistic fireball
model.  We also find that the intrinsic spectrum of the afterglow is
extincted both by dust (source frame $A_V \ga 1$ mag), and that the
shape of the extinction curve is  typical of young star-forming regions
like the Orion Nebula but is not typical of older star-forming or
starburst regions.  The $\approx$ 2 mag drop between the $R$ and the
$I$  bands can be explained by the far-ultraviolet non-linear component
of the extinction curve if $3 \la z \la 4$, and by the 2175 \AA\ bump
if $z \approx 2$; other redshifts are not consistent with the
observational data, given our general model.
\end{abstract}

\section{Introduction}

The GRB 970228 and GRB 980329 afterglows are among the most well
observed of the afterglows detected so far.  We use observations of
these afterglows to derive some of the properties of these bursts and
to constrain the nature of their host galaxies.

\section{GRB 970228 and the Nature of Its Host Galaxy}

We have determined the local galactic extinction toward the  GRB970228
field by comparing galaxy counts in two bands in this field to those in
the HDF, and by comparing the observed broad band colors of stars in
the GRB970228 field to the colors of library spectra of the same
spectral type.  We also estimate the extinction using the neutral
hydrogen column density and the amount of infrared dust emission toward
this field.  Combining the results of these methods, we find a best-fit
galactic extinction in the optical of $A_V=1.09^{+0.10}_{-0.20}$, which
implies a substantial dimming and change of the spectral slope of the
intrinsic GRB970228 afterglow.   Further details can be found in
Castander \& Lamb (1998, 1999a); see also Gonz\'alez et al. (1999).

\begin{figure}
\resizebox{\hsize}{!}{\includegraphics{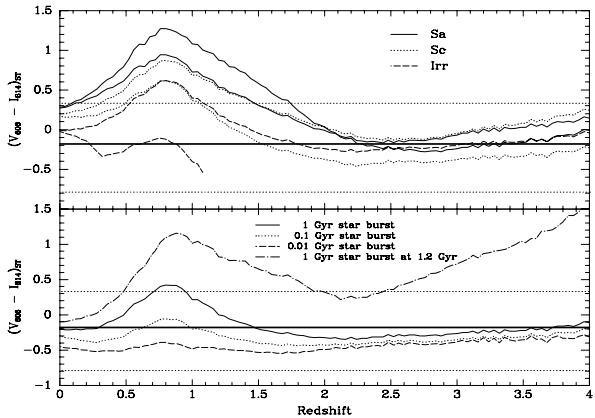}}
\caption
{WFPC2 colors for synthesized Spectral Energy Distributions (SEDs)
versus redshift.  The top panel shows the expected colors for spirals
and irregular galaxies including a local extinction of $A_V=1.09$.  The
thick lines include a K-correction, the thin lines include evolutionary
plus K-corrections.  The thick horizontal line is the best value
observed color and the horizontal dotted lines its $1\sigma$ error. 
The bottom panel shows models of 1 Gyr (solid line), 100 Myr (dotted
line) and 10 Myr (dashed line) starbursts observed at redshift $z$. 
The dot-dashed line represents a 1 Gyr starburst observed at redshift
$z$, 200 Myr after the cessation of star formation.  The cessation of
star formation, with the resulting disappearance of the most massive
stars, produces redder colors incompatible with the observations.}
\label{fig5}
\end{figure}

Re-analyzing the HST images, we measure a color $(V_{606}-I_{814})_{ST}
= -0.18^{+0.51}_{-0.61}$ for the extended source.  We constrain the
nature of the likely host galaxy of GRB 970228 by comparing this color
to those obtained from synthetic galaxy spectra computed with PEGASE
(\cite{FRV97}), taking into account the measured extinction.  The top
panel of Figure~\ref{fig5} shows the expected colors of an Sa, an Sc
and an Irregular galaxy with and without evolution included.  Galaxies
are only consistent with the observed color if they are at high
redshift ($z \gtrsim 1.0-1.5$), or have active star formation (like the
evolving Irr shown).

The bottom panel of Figure~\ref{fig5} better illustrates this point. It
shows that on-going bursts of star formation of duration shorter than 1
Gyr produce acceptable $V_{606}-I_{814}$ colors; longer duration bursts
are disfavored for redshifts $z \gtrsim 0.8$.  If we include the $H$
and $K$ magnitudes of Fruchter et al. (1998) in our analysis, our
conclusions are strengthened (see Castander \& Lamb 1999b for further
details). We conclude that the host galaxy must be undergoing star
formation, in agreement with our earlier result (Castander \& Lamb
1998; see also Castander \& Lamb 1999b).  If there is any extinction
present due to the host galaxy, this conclusion would be strengthened.

If the extended source is a galaxy with ongoing star formation, strong
emission lines are expected.  Tonry et al. (1997) and Kulkarni et al.
(1997) have tried to obtain the spectrum of the GRB970228 afterglow and
its associated nebulosity.  Neither observation revealed any obvious
emission lines.  The lack of observed [OII] and Ly${\alpha}$ emission
lines suggests that the redshift of the galaxy may lie in the range
$1.5 \lesssim z \lesssim 2.6$, considering the spectral coverage of the
observations.

\section{GRB 980329 and the Nature of Its Host Galaxy}


We model the observed radio through X-ray spectrum of the GRB 980329 
afterglow, and its evolution through time, as follows.  We take the
intrinsic spectrum to be a thrice broken power law, motivated by the
relativistic fireball model, in which spectral breaks may occur due to
synchrotron self-absorption, the synchrotron peak, and electron cooling
(see, e.g., Sari, Piran, \& Narayan 1998).  The spectrum that we fit is
a generalization of the spectrum expected in this model, in the sense
that we do not constrain the slopes of the four spectral segments, nor
the (power-law) rates at which these segments fade, a priori.

We allow the intrinsic spectrum to be modified in the following ways. 
First, we allow this spectrum to be extincted by dust and
absorbed by the Lyman limit at a single redshift, assumed to be the
redshift of the burst and its host galaxy.  We adopt the six parameter
ultraviolet extinction curve of Fitzpatrick \& Massa (1988) and the one
parameter optical and near-infrared extinction curve of Cardelli,
Clayton, \& Mathis (1989).
Finally, we redshift the modified spectrum to the observer's frame of
reference, and model the Lyman-$\alpha$ forest due to absorption by gas
clouds along the line-of-sight between the burst and the observer. 

\begin{figure}
\resizebox{\hsize}{!}{\includegraphics[trim=55 400 50 55]{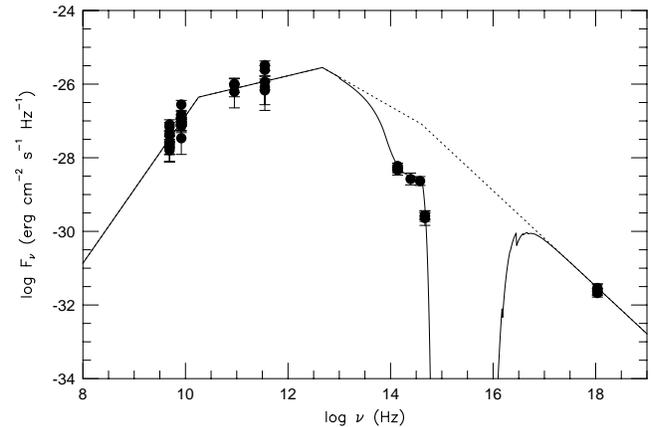}}
\caption
{The radio through X-ray spectrum of the afterglow of GRB 980329.  All
measurements have been scaled to a common time, approximately three
days after the GRB.  The solid curve is the best fit spectrum for an
isotropic fireball that expands into a homogeneous external medium,
extincted by dust at a redshift of  $z = 3.5$.  The dotted curve is the
un-extincted spectrum.
\label{romefig}}
\end{figure}

The afterglow of GRB 980329 is unique among afterglows observed to date
in that enough measurements of it have been taken to determine all the 
parameters of our model.  From the results of our fits, we draw six 
conclusions:  
(1) The inferred intrinsic spectrum of the afterglow is consistent with
the predictions of the simplest relativistic fireball model, in which
an isotropic fireball expands into a homogeneous external medium.   
(2) The intrinsic spectrum of the afterglow is extincted by dust
(source frame $A_V \ga 1$ mag).
(3) The linear component of the extinction curve is flat, which is 
typical of young star-forming regions like the Orion Nebula but is not
typical of older star-forming or starburst regions.
(4) The $\approx$ 2 mag drop between the $R$ and the $I$ bands can be
explained by the Ly-$\alpha$ forest if the burst redshift is $z \approx
5$ (Fruchter 1999), by the far-ultraviolet non-linear component of the
extinction curve if $3 \la z \la 5$, and by the 2175 \AA\ bump if $z
\approx 2$; other redshifts are not consistent with these data, given
this general model.  Djorgovski \etal (1999) report that $z < 3.9$
based upon the non-detection of the Ly-$\alpha$ forest in a Keck II
spectrum of the host galaxy.
(5) Assuming a redshift of $z = 3.5$ for illustrative purposes, using 
the observed breaks in the intrinsic spectrum, and solving for the 
physical properties of the fireball (see, e.g., Wijers \& Galama 1998),
we find that the energy of the fireball per unit solid angle is
$\cal{E}$ $\sim 10^{52}/4\pi$ erg sr$^{-1}$ if $\Omega_m = 0.3$ and
$\Omega_\Lambda = 0.7$.  
(6) Similarly, we find that the density of the external medium into 
which this fireball expands is $n \sim 10^{3}$ cm$^{-3}$ if $\Omega_m =
0.3$ and $\Omega_\Lambda = 0.7$.  This density suggests that GRB 980329
occurred in a molecular cloud, which is consistent with the fact that
the observed extinction features are characteristic of star-forming
regions.

\end{document}